# Nearfield Electromagnetic Effects on Einstein Special Relativity

William D. Walker
Norwegian University of Science and Technology (NTNU)
Previous research papers [1]
william.walker@vm.ntnu.no

**Abstract**

In this paper Maxwell equations are used to analyze the propagation of oscillating electric and magnetic fields from a moving electric dipole source. The results show that both the magnetic field and electric fields generated propagate faster than the speed of light in the nearfield and reduce to the speed of light as they propagate into the farfield of the source. In addition, the results show that the speed of the fields are dependant on the velocity of the source in the nearfield and only become independent in the farfield. These effects are shown to be the same whether the source or observation point is moving. Because these effects conflict with the assumptions on which Einstein's theory of special relativity theory is based, relativity theory is reanalyzed. The analysis shows that the relativistic gamma factor is dependant on whether the analysis is performed using nearfield or farfield propagating EM fields. In the nearfield, gamma is approximately one indicating that the coordinate transforms are Galilean in the nearfield. In the farfield the gamma factor reduces to the standard known relativistic formula indicating that they are approximately valid in the farfield. Because time dilation and space contraction depend on whether nearfield or farfield propagating fields are used in their analysis, it is proposed that Einstein relativistic effects are an illusion created by the propagating EM fields used in their measurement. Instead space and time are proposed to not be flexible as indicated by Galilean relativity.

**Dipole analysis**

Maxwell equations will be used to analyze the propagation electromagnetic (EM) fields generated by an electric dipole source. The results will show that the fields propagate faster than light in the nearfield and reduce to the speed of light as they propagate into the farfield of the source [ref. authors papers - 2, 3, 4, 5, 6, 7, 8, 9, 10 and other authors papers - 11, 12]. In addition, the fields will be shown to be velocity dependant in the nearfield and only independent in the farfield. These effects will be shown to be the same whether the source or observation point is moving.

Dipole moving toward stationary observer

In the first case assume an electric dipole (p) of length ($r'$) is moving (v) toward a stationary observer (O). From the figure (Fig. 1) it can be seen that the distance (R) from the dipole to the observation point, along the axis of propagation, is approximately: $R = r - vt$, where the distance from the origin of the reference coordinate is (r) and the angle ($\theta$) is 90 deg.

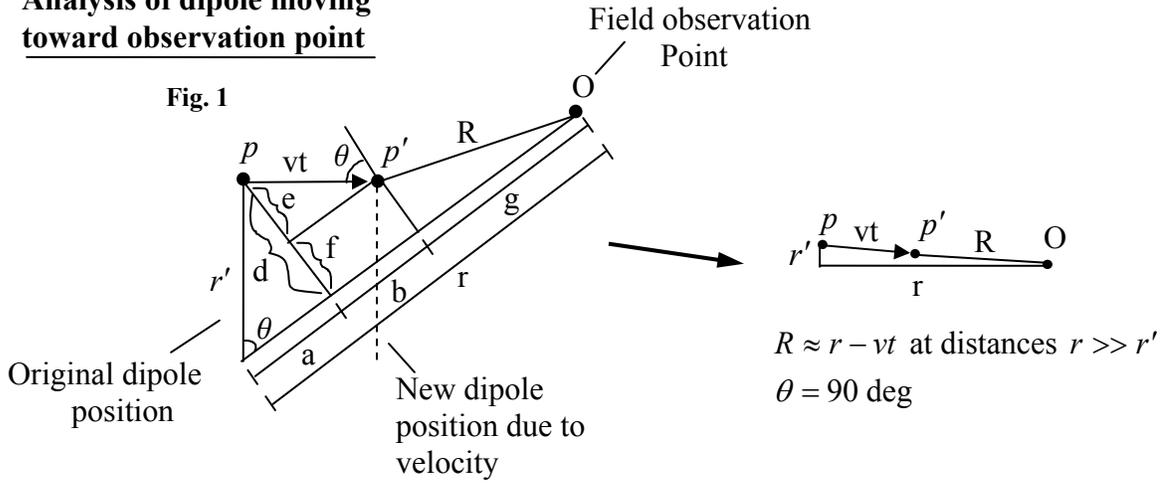

**Analysis of dipole moving toward observation point**

Fig. 1

Original dipole position

New dipole position due to velocity

Field observation Point

$R \approx r - vt$ at distances $r >> r'$

$\theta = 90$ deg

From the geometry in the above figure (Fig. 1) one can determine the distance (R) between the moving dipole source (p) and the observation point (O).

$$R = \sqrt{f^2 + g^2} \quad g = r - a - b \quad a = r'Cos(\theta) \quad d = r'Sin(\theta)$$
$$e = vt\, Sin(90 - \theta) = vt\, Cos(\theta)$$
$$f = d - e = r'\, Sin(\theta) - vt\, Cos(\theta) \quad b = vt\, Cos(90 - \theta) = Sin(\theta)$$

$$\therefore R = r\sqrt{\left[\xi Sin(\theta) - \frac{vt}{r}Cos(\theta)\right]^2 + \left[1 - \xi Cos(\theta) - \frac{vt}{r}Sin(\theta)\right]^2} \quad \text{where } \xi = \frac{r'}{r} \quad (1)$$

It should be noted that if the dipole moves away from the observation point then the dipole velocity v becomes –v in the above relation for R. At observation distances (r) much larger than the a dipole length $(r')$ and for an observation point along the axis of motion $\theta = 90$ deg, the above relation simplifies to:

$$R = r - vt \quad (2)$$

Observer moving toward stationary dipole

In the next case, assume that the observation point (O) is moving (v) toward a stationary dipole (p) of length $(r')$. From the figure (Fig. 2) below it can be seen that the distance (R) from the dipole to the observation point, along the axis of propagation, is approximately: $R = r - vt$, where the distance from the origin of the reference coordinate is (r) and the angle $(\theta)$ is 90 deg.

**Analysis of observation
point moving toward dipole**

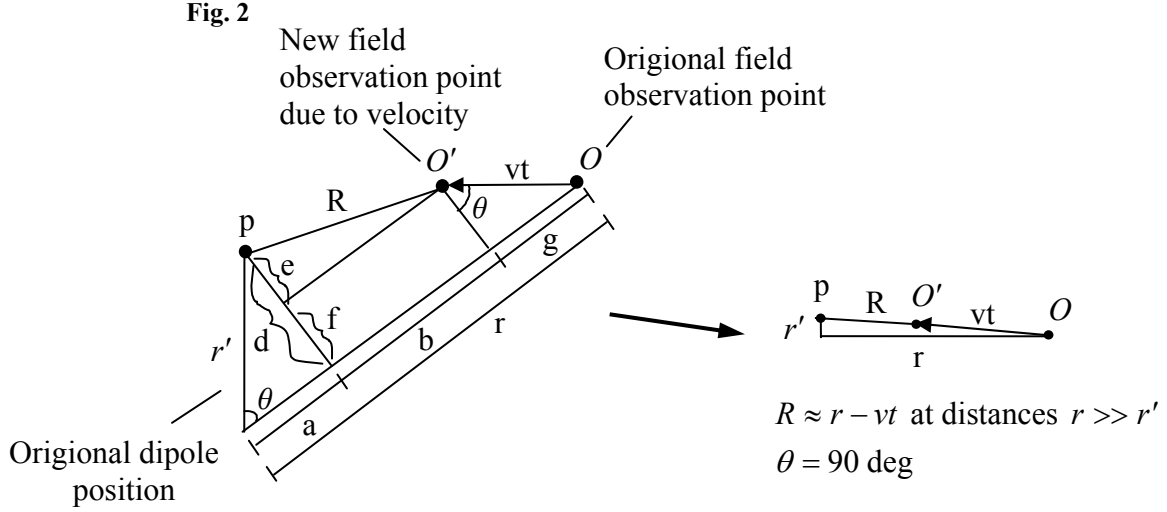

Fig. 2

$R = \sqrt{e^2 + b^2}$   $g = r - a - b$   $\therefore b = r - a - g$   $a = r' Cos(\theta)$
$d = r' Sin(\theta)$   $g = vt\, Sin(\theta)$   $f = vt\, Cos(\theta)$
$e = d - f = r' Sin(\theta) - vt\, Cos(\theta)$

$$\therefore R = r\sqrt{\left[\xi Sin(\theta) - \frac{vt}{r} Cos(\theta)\right]^2 + \left[1 - \xi Cos(\theta) - \frac{vt}{r} Sin(\theta)\right]^2} \quad \text{where } \xi = \frac{r'}{r} \quad (3)$$

Again it should be noted that if the dipole moves away from the observation point then the dipole velocity v becomes −v in the above relation for R. At observation distances (r) much larger than the a dipole length $(r')$ and for an observation point along the axis of motion $\theta = 90$ deg, the above relation simplifies to:

$$R = r - vt \quad (4)$$

These two analysis clearly show that the distance between the dipole and observation point is the same independent of whether the dipole or observation point is moving, which is consistent with both Galilean and Einstein special relativity.

Magnetic field propagation analysis

To calculate the magnetic field (B) one can insert the above expression for R into the retarded vector potential (A) [2]:

$$A = \frac{1}{c}\int \frac{J(t-\frac{R}{c})}{R} dv' \tag{5}$$

$$\text{where:} \quad J(t) = \frac{dp}{dt}\hat{z} \quad p = p_o e^{-i\omega t} \quad \hat{z} = Cos(\theta)\hat{r} - Sin(\theta)\hat{\theta} \tag{6}$$

The B field can then be calculated by taking the curl of the vector potential:

$$B = \nabla \times A \tag{7}$$

yielding (ref. Fig. 3):

$$B_\phi = \frac{\omega p_o}{c} \frac{e^{i\left\{kr - \omega\left[1+\frac{v}{c}\right]t\right\}}}{(r-vt)^2}\left[-k(r-vt)-i\right] \tag{8}$$

**Fig. 3** Maple 7 program code used to derive above result (Eq. 8)
```
with(linalg):
H := [r, theta, phi];
R:=r-v*t;
f:=exp^(I*(k*R))/R;
p(t):=Po*exp(-I*w*t);
s:=f/c*diff(p(t),t);
A:=[s*cos(theta),-s*sin(theta),0];
B:=curl(A,H,coords=spherical);
theta:=Pi/2;
Bphi:=simplify(B[3]);
```

Combining the real and imaginary terms in the brackets into a phasor ( $Mag\ e^{i(ph)}$ ) with magnitude (Mag) and phase (ph) the B field becomes:

$$B_\phi = \frac{\omega p_o}{c} Mag \frac{e^{i\left\{ph - \omega\left[1+\frac{v}{c}\right]t\right\}}}{(r-vt)^2} \tag{9}$$

$$\text{where} \quad ph = \left[kr - Cos^{-1}\left(\frac{x}{\sqrt{x^2+y^2}}\right)\right], \quad x = -k(r-vt), \quad y = -1 \tag{10}$$

The phase speed of the B field can then be calculated using [13]:

$$c_{ph} = c_o k \Big/ \frac{\partial \theta}{\partial r} = c_o \Big/ \frac{\partial \theta}{\partial (kr)} \qquad (11)$$

Note that the formula: $c_{ph} = \omega/k$ formula is not valid when the phase is nonlinear (ref. Eq. 10 and Eq. 55 in reference [2]). The phase speed for the magnetic field is then calculated to be:

$$c_{ph} = c_o \left\{ 1 + \frac{1}{(kr)^2 \left[1 - \dfrac{v}{c}\right]} \right\} \qquad (12)$$

**Fig. 4**                      **Fig. 5**

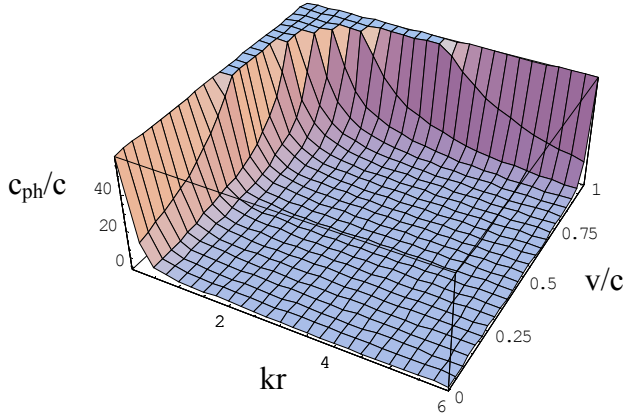
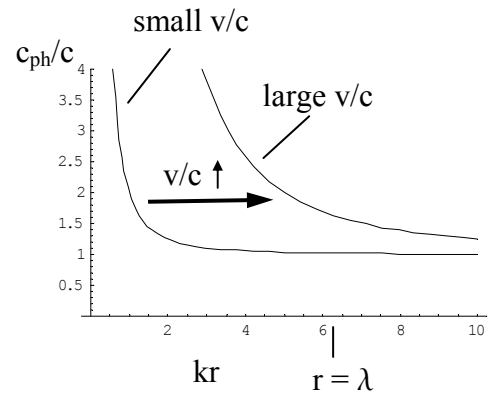

Contrary to expectation (as discussed in the Michelson-Moreley type experiments), this solution shows (ref. Eq. 12, Fig. 4, 5) that the phase speed of the magnetic field emitted by an electric dipole is velocity dependent in the nearfield ($kr < 2\pi$. i.e $r < \lambda$) and only approximately velocity independent and constant in the farfield ($r > \lambda$).

Electric field propagation analysis

Similar analysis using the longitudinal electric field generated by an electric dipole source yields the same phase speed relation as derived for the B field [2] consequently this would yield a similar gamma function. But, analysis of the transverse electric field generated by a dipole source is very different, showing that the field is created about one quarter wavelength outside the source, generating field components which propagate superluminally both toward and away from the source. As the transverse field propagates into the farfield, the phase speed reduces to the speed of light. Since the gamma function analysis would be more complicated for this field it is not included in this discussion. But it is noted that the resultant gamma function should behave similarly near the creation point where the field is instantaneous and in the farfield where the field propagates at the speed of light.

**Special relativistic consequences**

Since special relativity theory [14] is based on the premise that the speed of light is constant and velocity independent, the above results suggest that relativity theory may be different when near-field propagating fields are used to analyze moving systems. When far-field propagating fields are used normal relativity theory can be used.

Coordinate frames

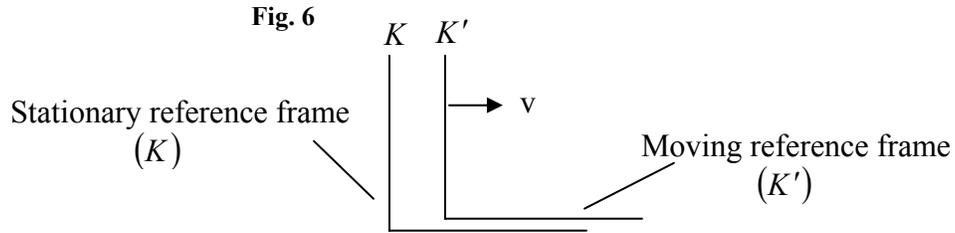

Fig. 6

Stationary reference frame $(K)$

Moving reference frame $(K')$

Galilean transformation

$$\boxed{\begin{aligned} x' &= (x - vt) \\ x &= (x' + vt') \\ \text{where } t &= t' \end{aligned}}\quad \text{solving for x yields:} \tag{13}$$
(14)
(15)

Near-field relativity

As in the normal derivation of the Lorentz transformations, assume the reciprocal character of these equations:

$$\boxed{\begin{aligned} x' &= \gamma(x - vt) \\ x &= \gamma(x' + vt') \end{aligned}} \quad \text{where the function } \gamma \text{ is to be determined} \tag{16}$$
(17)

Taking the derivative of the first equation above with respect to x yields:

$$\frac{\partial x'}{\partial x} = \gamma\left(1 - v\frac{\partial t}{\partial x}\right) \qquad \text{Note } \gamma \text{ assumed to be independent of } x \tag{18}$$

Taking the derivative of the second equation above with respect to $x'$ yields:

$$\frac{\partial x}{\partial x'} = \gamma\left(1 + v\frac{\partial t'}{\partial x'}\right) \qquad \text{Note } \gamma \text{ assumed to be independent of } x' \tag{19}$$

Taking the inverse of the second to last equation and equating it to the above equation yields:

$$\frac{\partial x}{\partial x'} = \frac{1}{\gamma\left(1 - v\dfrac{\partial t}{\partial x}\right)} = \gamma\left(1 + v\frac{\partial t'}{\partial x'}\right) \tag{20}$$

Assume a B field is transmitted from a dipole in the stationary frame and is observed in the two reference frames. According to the stationary reference frame $(K)$ the speed of the propagating B field is observed to be (ref. Eq. 12, where v = 0):

$$c_{ph} = \frac{\partial x}{\partial t} = c_o \left\{ 1 + \frac{1}{(kr)^2} \right\} \tag{21}$$

But according to the moving reference frame $(K')$ the speed of the propagating B field is observed to be (ref. Eq. 12):

$$c'_{ph} = \frac{\partial x'}{\partial t'} = c_o \left\{ 1 + \frac{1}{(kr)^2 \left[ 1 - \frac{v}{c} \right]} \right\} \tag{22}$$

Inserting these phase speed relations into the second to last equation (ref. Eq. 20) yields:

$$\frac{1}{\gamma \left( 1 - \frac{v}{c_{ph}} \right)} = \gamma \left( 1 + \frac{v}{c'_{ph}} \right) \tag{23}$$

Solving for $\gamma$ yields:

$$\gamma = \frac{1}{\sqrt{\left(1 - \frac{v}{c_{ph}}\right)\left(1 + \frac{v}{c'_{ph}}\right)}} \tag{24}$$

$$\therefore \gamma = \frac{1}{\sqrt{\left(1 + \frac{\frac{v}{c}}{1 + \frac{1}{(kr)^2}}\right)\left(1 + \frac{\frac{v}{c}}{1 + \frac{1}{(kr)^2\left[1-\frac{v}{c}\right]}}\right)}} \quad \underset{r \ll \lambda}{=} 1 \quad \underset{r \gg \lambda}{=} \frac{1}{\sqrt{\left(1 + \left(\frac{v}{c}\right)^2\right)}} \tag{25}$$

reduces to one in nearfield

reduces to Einstein relativity gamma function in farfield

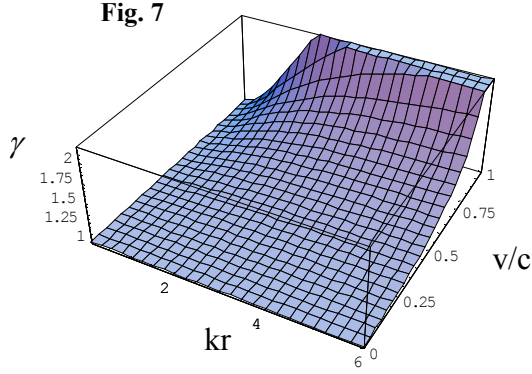

**Fig. 7**

**Fig. 8**

**Fig. 9**

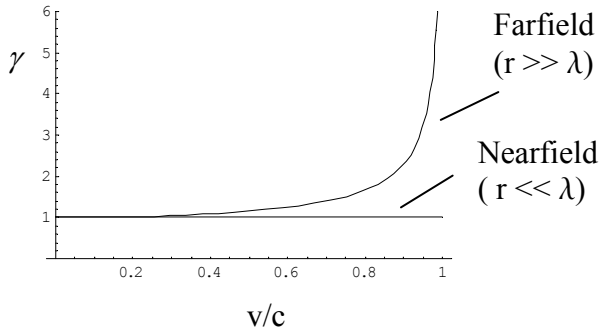

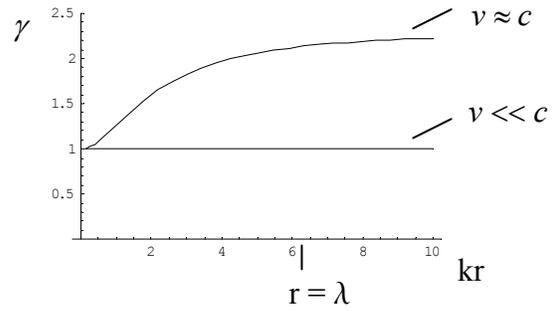

Note that $\gamma$ (ref. Eq. 25, Fig. 7, 8, 9) is approximately independent of r in the farfield ($r > \lambda$) and also in the nearfield ($r \ll \lambda$), specially when $v \ll c$, thus justifying for these two limits the premise made at the beginning of the analysis that $\gamma$ is independent space (ref. Eq. 18, 19).

Time transform derivation
Substituting (ref. Eq. 16): $x' = \gamma(x - vt)$ into the relation (ref. Eq. 17): $x = \gamma(x' + vt')$ and solving for $t'$ yields the near-field time transform relative the moving observer:

$$\boxed{t' = \gamma\left(t - \frac{x}{v}\left\{\frac{1}{\gamma^2} - 1\right\}\right)} \quad \begin{aligned} &\underset{r \ll \lambda}{=} \gamma t \quad\quad\text{— get Galilean relativity time transform in nearfield} \\ &\underset{r \gg \lambda}{=} \gamma\left(t - \frac{v}{c^2}x\right)\text{— get Einstein relativity time transform in farfield} \end{aligned} \quad (26)$$

where $\gamma$ is the function derived above

Substituting (ref. Eq. 17): $x = \gamma(x' + vt')$ into the relation (ref. Eq. 16): $x' = \gamma(x - vt)$ and solving for t yields the near-field time transform relative to the stationary observer:

$$\boxed{t = \gamma\left(t' + \frac{x'}{v}\left\{\frac{1}{\gamma^2} - 1\right\}\right)} \quad \begin{aligned} &\underset{r \ll \lambda}{=} \gamma t' \quad\quad\text{— get Galilean relativity time transform in nearfield} \\ &\underset{r \gg \lambda}{=} \gamma\left(t' - \frac{v}{c^2}x'\right)\text{— get Einstein relativity time transform in farfield} \end{aligned} \quad (27)$$

where $\gamma$ is the function derived above

Length contraction
A rod of length $\Delta l$ in a moving frame $(K)$ will appear contracted by an observer in a stationary frame $(K')$:

$$\Delta l' = x_2' - x_1' = \gamma(x_2 - vt_2) - \gamma(x_1 - vt_1)$$

Since time is the same in the stationary frame: $t_2 = t_1$

$$\boxed{\therefore \Delta l' = \gamma \Delta l} \quad \text{where } \Delta l = x_2 - x_1 \tag{28}$$

solving the above relation for $\Delta l'$ yields:

$$\boxed{\therefore \Delta l = \frac{\Delta l'}{\gamma}} \quad \text{i.e. stationary observer sees rod contracted} \tag{29}$$

Note that observers using a near-field propagating magnetic field will see no length contraction since $\gamma = 1$ in the nearfield, whereas observers using a far-field propagating magnetic field will see the usual Lorentz contraction.

Time dilation
A clock in a moving frame will appear to run slower by an observer in a stationary frame:

$$\Delta t' = t_2' - t_1' = \gamma\left(t_2 - \frac{x_2}{v}\left\{\frac{1}{\gamma^2} - 1\right\}\right) - \gamma\left(t_1 - \frac{x_1}{v}\left\{\frac{1}{\gamma^2} - 1\right\}\right) \tag{30}$$

Since the clock is at the same position in the stationary frame: $x_2 = x_1$

$$\boxed{\therefore \Delta t' = \gamma \Delta t} \quad \text{where } \Delta t = t_2 - t_1 \tag{31}$$

i.e. time interval observed in a moving frame appears larger
than the time interval in a stationary frame.

Solving the above relation for $\Delta t'$ yields:

$$\boxed{\Delta t = \frac{\Delta t'}{\gamma}} \tag{32}$$

Note that observers using a near-field propagating magnetic field will see no time dilation since $\gamma = 1$ in the nearfield, whereas observers using a far-field propagating magnetic field will see the usual relativistic time dilation.

Velocity transform

Using Eq. 20: $\dfrac{1}{\gamma\left(1+\dfrac{v}{\dot{x}'}\right)} = \gamma\left(1-\dfrac{v}{\dot{x}}\right)$ , where $\gamma$ is a function of kr as previously derived,

and solving for velocity $(\dot{x})$ yields:

$$\boxed{\dot{x} = \dfrac{v\left(1+\dfrac{v}{\dot{x}'}\right)}{1+\dfrac{v}{\dot{x}'}-\dfrac{1}{\gamma^2}}} \quad \underset{r \ll \lambda}{=} \dot{x}'+v \quad \text{get Galilean relativity velocity transform in nearfield (i.e. time derivative of Eq. 14)}$$

$$\underset{r \gg \lambda}{=} \dfrac{\dot{x}'+v}{1+\dfrac{v}{c^2}\dot{x}'} \quad \text{get Einstein relativity velocity transform in farfield} \quad (33)$$

Nearfield
In the nearfield the velocity transform reduces to the Galilean velocity transform where the v = infinity is invariant [*i.e. get* $\dot{x} = \infty$ *for* $(\dot{x}' = \infty$ *or* $v = \infty)$]. Also note that in the nearfield superluminal signals are permitted (*i.e. can get* $\dot{x} > c$ *for* $(\dot{x}' < c$ *and* $v < c))$.

Farfield
In the farfield the velocity transform reduces to the normal relativistic velocity transform where the v = c is invariant [*i.e. get* $\dot{x} = c$ *for* $(\dot{x}' = c$ *or* $v = c)$]. Also note that in the farfield superluminal signals are not permitted (*i.e. get* $\dot{x} < c$ *for* $(\dot{x}' < c$ *and* $v < c))$.

Similarly the inverse transform can be obtained by solving (Eq. 20) for velocity $(\dot{x}')$ yielding:

$$\boxed{\dot{x}' = \dfrac{v\left(1-\dfrac{v}{\dot{x}}\right)}{\dfrac{1}{\gamma^2}+\dfrac{v}{\dot{x}'}-1}} \quad \underset{r \ll \lambda}{=} \dot{x}-v \quad \text{get Galilean relativity velocity transform in nearfield (i.e. time derivative of Eq. 13)}$$

$$\underset{r \gg \lambda}{=} \dfrac{\dot{x}-v}{1-\dfrac{v}{c^2}\dot{x}} \quad \text{get Einstein relativity velocity transform in farfield} \quad (34)$$

Doppler Shift
The magnetic field observed by a moving observer (moving toward the source) has been calculated to be (ref. Eq. 9):

$$B_\phi = \dfrac{\omega p_o}{c} Mag \dfrac{e^{i\left\{ph-\omega\left[1+\frac{v}{c}\right]t\right\}}}{(r-vt)^2}$$

where the frequency observed by the moving observer is: $\omega' = \omega\left(1+\dfrac{v}{c}\right)$

Inserting the Lorentz transform for time (ref. Eq. 27): $t = \gamma\left(t' + \frac{x'}{v}\left\{\frac{1}{\gamma^2} - 1\right\}\right)$, where $x' = 0$ (corresponding to observer at the origin of the moving frame) yields:

$$\boxed{\frac{f'}{f} = \gamma\left(1 + \frac{v}{c}\right)} \quad \begin{array}{l} \underset{r \ll \lambda}{=} 1 + \frac{v}{c} \quad \text{get Galilean Doppler shift in nearfield} \\[1em] \underset{r \gg \lambda}{=} \frac{\sqrt{1 + \frac{v}{c}}}{\sqrt{1 - \frac{v}{c}}} \quad \text{get Einstein relativity Doppler shift in farfield} \end{array} \quad (35)$$

In the nearfield moving observers will see the Galilean Doppler shift since $\gamma = 1$ (independent of velocity), whereas in the farfield moving observers will see the usual relativistic Doppler shift.

**Discussion**

It has been shown in the previous analysis that the space and time coordinate transforms reduce to the Galilean relativity transformations in the nearfield and in the farfield they reduce to the Einstein relativity transformations. Intuitively this can be understood because in the nearfield the propagation speed of EM fields are approximately infinite. Substitution of ($c = \infty$) into the Einstein relativity (Lorentz) transforms yields the Galilean transformations:

$$x' = \frac{(x - vt)}{\sqrt{1 + \left(\frac{v}{c}\right)^2}} \underset{c = \infty}{=} x - vt \qquad t' = \frac{\left(t - \frac{vx}{c^2}\right)}{\sqrt{1 + \left(\frac{v}{c}\right)^2}} \underset{c = \infty}{=} t$$

$$x = \frac{(x' + vt')}{\sqrt{1 + \left(\frac{v}{c}\right)^2}} \underset{c = \infty}{=} x' + vt' \qquad t = \frac{\left(t' + \frac{vx'}{c^2}\right)}{\sqrt{1 + \left(\frac{v}{c}\right)^2}} \underset{c = \infty}{=} t'$$

It is apparent from all the previous analysis that the space and time transformations seem to depend on whether near-field or far-field propagating fields are used. But according to Einstein special relativity the effects of space contraction and time dilation should be real and independent of near-field or far-field behavior of propagating EM fields. To resolve this dilemma it is proposed that the Einstein relativity transformations are an illusion caused by the propagation delays of EM fields used in the measurement of time and space. Instead it is proposed that the real space and time transformations are Galilean and only appear to be different when propagating far-field EM fields are used to in their measurement. If near-field propagating EM fields are used then the propagation times delays are nearly zero and do not affect the transformations.

This is exemplified in Einstein relativity where observers in inertial frames measure the same space and time measurements but see space contraction and time dilation effects in other different inertial frames. Although space and time are correctly measured by observers in their own inertial frame using propagating EM fields, they create artificial distortions when they are used to measure space and time in other different inertial frames.

Near-field gamma function

In the derivation of the near-field relativistic transforms it was assumed that the gamma function is independent of space (Eq. 18, 19), which is true in the farfield ($r > \lambda$) and also very near the source ($r \ll \lambda$). In between theses two regions the approximation breaks down slightly and will cause the gamma function to alter slightly. To improve the modeling of the gamma function the resultant equation (Eq. 25) can be repeatedly inserted in (Eq. 18, 19) and then calculated again. Because the purpose of this paper is to demonstrate that relativity theory is only an illusion caused by the time delays of EM fields used in the measurement of relativistic effects, it is sufficient to look only at the gamma function very near the source and very far from the source, where the derived gamma function is approximately valid. In these two extreme limits it has been shown that Einstein relativity is only valid in the farfield and reduces to Galilean relativity in the nearfield, very near the source.

Phase speed vs. Group speed

The previous derivation of the nearfield relativistic transforms assumed that monochromatic fields would be used in the measurement space of and time. If the source is not monochromatic then the arguments would have to redone using group speed. It can be shown that for narrow banded ($\Delta f / f_o \geq 1/100$) sources, the group speed, particularly in the nearfield, is very similar to the phase speed [2] yielding a near-field relativistic gamma function qualitatively very similar to what was derived for the EM phase speed. In the above referenced paper it has been shown that dispersion is nonlinear in the nearfield of a dipole and only linear in the farfield. But provided the field is narrow banded then the dispersion is nearly constant over the bandwidth of the field causing the field to propagate nearly undistorted at the group speed. It should be noted that because pulses can be very broadband they may distort significantly as the propagate in the nearfield, but the distortion process would slow down as the pulse propagates into the farfield. Therefore if pulses are to be used in an experiment, they should be narrow banded

Experiments

The reason many of the effects derived in this paper have not been experimentally observed is that most experiments measure relativistic effects using farfield EM fields. Using light sources, these effects must be measured using source to detector distances within a wavelength of light, which is on the order of atomic distances (nanometers). Because experiments of this nature are difficult to do, larger wavelength (~100MHz) EM near-field propagating fields should be used. For instance experiments conducted by the author using simple dipole antennas (400MHz) have demonstrated the superluminal behavior of near-field transverse electric fields [2].

**Conclusion**

The analysis in this paper has shown that, according to Maxwell's equations the propagation speed of EM fields are nearly infinite in the nearfield and reduces to the speed of light in the farfield. In addition, the propagation speed is dependant on the source in the nearfield and only approximately independent in the farfield. The field propagation has been shown to be independent of whether the source or observation point is moving, which is consistent with Galilean relativity. Einstein relativity is also based on this assumption but in addition assumes that the speed of light is constant. Since this assumption has been shown to not to be valid in the nearfield, relativity theory has been reanalyzed and has been shown in this paper to reduce to Galilean relativity in the nearfield and to approximately Einstein relativity in the farfield. Because the absolute nature of space and time can not depend on experimental configuration (i.e. use of near-field or far-field EM fields), it is proposed that that Einstein relativistic effects are an illusion caused by the propagation time delays of EM fields used to measure time and space. Instead Galilean relativity is proposed to describe the real nature of space and time. The Lorentz transforms, with the new proposed definition of gamma is still very useful for calculating apparent space contraction and time dilation effects when high velocity systems are observed from non-moving reference frames. If a physical system is measured using near-field propagating EM fields then the apparent space-time changes will not be very noticeable and simple Galilean transformations can be used, but if physical systems are measured with far-field EM fields then the Lorentz transform can be used to calculated the altered space-time illusion, but it must be stressed that these effects are not real. Time and space are not to be interpreted as being flexible.

"If the speed of light is the least bit affected by the speed of the light source, then my whole theory of relativity and theory of gravity is false. " – Albert E. Einstein